\documentclass[prl,aps,twocolumn,showpacs]{revtex4}
\usepackage{graphicx}
\usepackage{amssymb}
\usepackage{subfigure}
\begin{document}
\title{Fluctuations and Dispersal Rates in Population Dyanmics}
\author{David A. Kessler$^{1}$ \& Leonard M. Sander$^{2}$}
\affiliation{$^1$Department of Physics, Bar-Ilan University, Ramat-Gan, IL52900, Israel\\
 $^2$Department of Physics \& Michigan Center for Theoretical Physics, University of Michigan, Ann Arbor, MI 48109 USA}
\begin{abstract}
Dispersal of species to find a more favorable habitat is important in population dynamics. 
Dispersal rates  evolve in response to the relative success of different dispersal strategies.  In a simplified deterministic treatment (J. Dockery, V. Hutson, K. Mischaikow, et al., J.  Math. Bio.  {\bf 37}, 61 (1998)) of two species which differ only in their dispersal rates the slow species \emph{always} dominates.  We
demonstrate that fluctuations can change this conclusion and can lead to dominance
by the fast species or to coexistence, depending on parameters. We discuss two different effects of fluctuations, and show that our results are consistent with more complex treatments that find that selected dispersal rates are not monotonic with the cost of migration.
\end{abstract}
\pacs{87.23.Cc, 87.10Mn} \maketitle

Dispersal plays an essential role in the dynamics of populations. This effect is
particularly important when different parts of a habitat are more or less desirable because of a distribution of resources, varying temperatures, etc.
Rates of dispersal can evolve in response to such distributions \citep{Johnson}.  In this paper we try to understand this evolution by considering the competition of two species,``fast'' (F) and  ``slow'' (S),  that differ only by their dispersal rate. We ask who ``wins'', i.e. whether one species drives the other to extinction in the long-time limit. We also consider the possibility of  coexistence.

There are two contradictory guesses that one might make. On the one hand, if a species is well adapted to part of a habitat, it would be a waste of resources to wander away immediately; this is exactly what  species F does. We might postulate that F  will ultimately be displaced by S provided there is some probability for S to find all the parts of the system. This idea is supported by the work  of Hastings \citep{Hastings} and 
Dockery, et al. \citep{Dockery} who formulated this problem in terms of a pair of deterministic reaction-diffusion equations.  They proved that S always drives F to extinction for any non-uniform habitat. 

On the other hand Hamilton and May \cite{HM} considered a model in which F may be preferred. In this picture there are discrete sites that allow one organism per site. In each generation each organism produces offspring that can either remain at the site  or spread at random to all other sites while paying a cost of migration. The adult  in the next generation is chosen at random among the young present at the site. It is easy to see that S does not always win in this model. For example, if S has zero dispersal rate  then S will go extinct.  This results from the fact that there is a finite probability that a young F will take over any site even if it is outnumbered by S, and S can never return. This result depends on the stochastic nature of the model. Later work generalized the model to have more than one adult per site \cite{CHM}. In this model  there is an optimal dispersal rate which depends on the cost of migration.

We will try to understand the relationship between the two results above by using an agent based model  which is closely related to the reaction-diffusion equations of \cite{Dockery}. Of course, for very large populations the deterministic model must be correct. We explore the role of fluctuations for moderate populations. 

We are motivated to do this by the fact  that stochastic effects should be important in real populations in a  patchy environment. Further, it is known in other contexts  \citep{Witten,Derrida,KSN} that
reaction-diffusion equations  can give misleading results. We will find that there are two effects that lead to differences from the deterministic approach. For a patchy environment  discreteness of individuals means that migration across unfavorable regions is much less frequent  than that predicted by the differential equations; this promotes coexistence. For nearly uniform resources there is another, more subtle, effect: fluctuations in population density increase death rates so that F, whose populations are more uniform, is favored. To our knowledge this mechanism has not been discussed previously.  Our simple picture explicates the essential features of more complex treatments based on data for real populations \citep{Hanski}.

In our agent-based model the two species live at sites on a one-dimensional lattice with lattice constant $h$. Positions are given by $x_j=(j-1/2)h$, $j=1\ldots N_x$. (We can think of $h$ as being comparable to the minimum,length scale over which the environment varies.)  The numbers of  agents at site $j$ are $F(x_j,t), S(x_j,t)$. The agents are identical except for their rates to jump to adjacent sites: $D_f > D_s$. The birth rates per agent of either type are non-uniform to reflect the non-uniform habitat, and are given by a specified function of position $\alpha(x_j)$.  To limit populations we assume the death rate is population dependent;  we use the logistic model, so that the death rate per agent is $\beta (F+S)$; competition is encoded in the death rate. Time is in units of the inverse growth rate for the population in the most favorable region of the habitat.
If F drives S to extinction, we take this as an indication that in a natural population the dispersal rate will evolve upwards, and \emph{vice versa}. 

The  algorithm for our agent-based model is quite straightforward. First, we pick a time step $dt$.   At each point there is a number of individuals of each type at each site. We take the probability to move left or right to be  $D_{f} dt$ for the fast species and $D_s dt$ for the slow. In most of the simulations we take $D_f=1, D_s=0.25$. The number of agents that move in each direction is given by the binomial distribution. Similarly, the probability of birth per agent  is $\alpha dt$. The number of births is a binomial deviate; this number is added to the original number at the site for each species. For deaths the corresponding probability for each agent is $\beta(F+S)dt$.  

We adopt a simple landscape for $\alpha(x_j)$ which has two `oases' and a `desert' between them: $\alpha(x)= 1+\eta \cos(2\pi x/L)$  where $L=N_x h$ is the length of the system: see Figure 1(a).  The parameter $\eta$ gives the range of variation of the birth rate. If $\eta=0$ the birth rate is the same everywhere;  large $\eta$ indicates a patchy landscape and a desert which imposes a large cost of migration. 
We take $\beta=1/N_o$ so that in the absence of diffusion, and for one species,  the population at each site saturates at the carrying capacity:  $\alpha/\beta = N_o (1+\eta \cos( 2\pi x/L))$. The initial condition is that S is localized on the left oasis, and F on the right, and the boundary condition is that no flux passes through $x=0$ or $x=L$.

The result of many simulations of our model in this landscape is shown in Figure 1(b). The striking feature  is that at small population densities F always wins.  Further,  for large $\eta$  and relatively small $N_o$  extinction takes a very long time so that in a real situation there would appear to be stable coexistence. Thus the region where F persists for very long times is \emph{non-monotonic}.\begin{figure}
\subfigure[ ]{\includegraphics[width=3in]{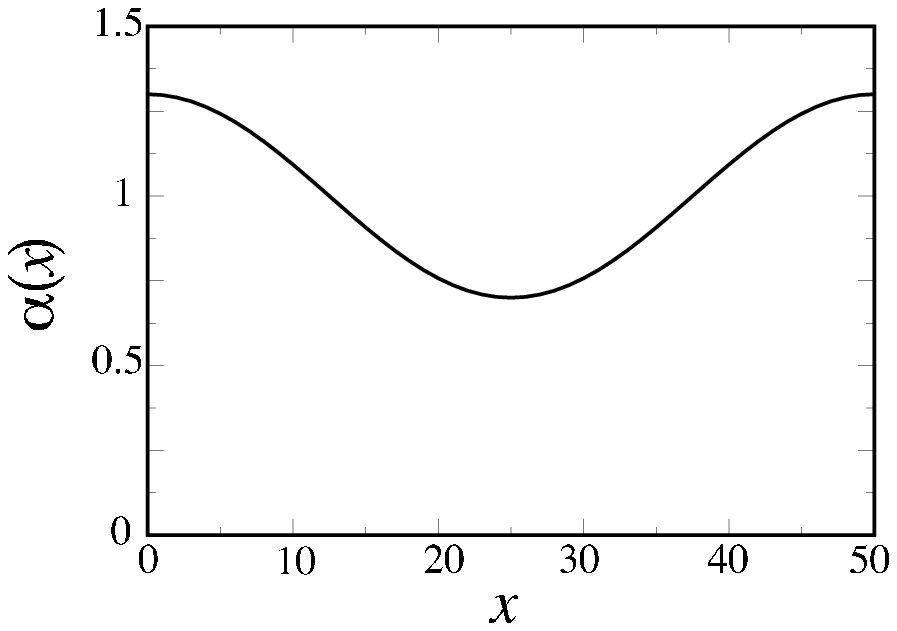}}
\subfigure[ ]{\includegraphics[width=3in]{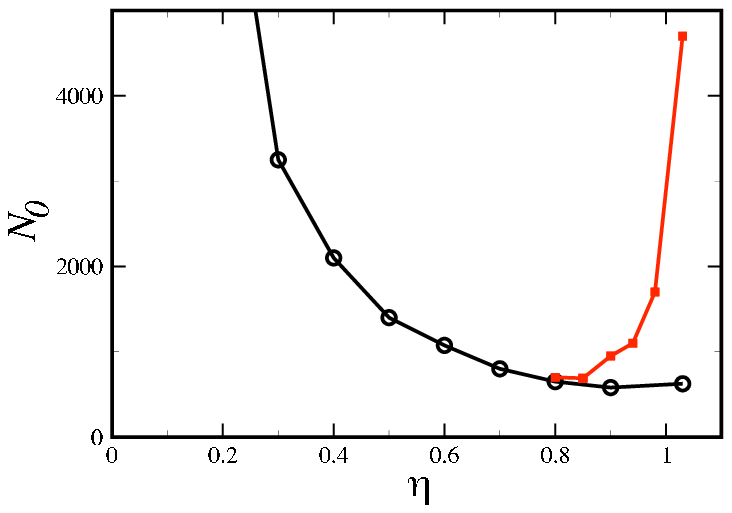}}
\caption{ (a). The landscape for a system of length 50.  (b). Simulations results from the agent-based model; $D_f=1, D_s=0.25$. Above the line marked with circles  ($\circ$) S wins, and below, F. For large $\eta$ there is  long-lived coexistence: below the line marked with squares ($\square$)  the extinction of F takes more than 10 times the  prediction of Eq. (1) to fall below one individual per cell.}
\end{figure}

The reaction-diffusion equation version of the model gives different results \citep{Hastings,Dockery}. The equations are:
\begin{eqnarray}
 \frac{\partial  f}{\partial t} &=& \tilde D_f \frac{\partial^2 f}{\partial x^2} + \alpha f
        -\tilde \beta  f( f+ s),  \nonumber \\
\frac{\partial s}{\partial t} &=& \tilde D_f \frac{\partial^2 s}{\partial x^2} + \alpha s
        -\tilde \beta s( f+  s).
        \label{RD}
   \end{eqnarray}
Here the parameters of the spatially discrete individual-based model are replaced by continuum quantities: $ f = F/h,  s=S/h, \tilde D_{f,s} = D_{f,s} h^2, \tilde \beta = \beta h$.  We should expect these equations to give the same results as the individual-based model for large $N_o$  and sufficiently smooth $\alpha(x)$. 
 
The results of \citep{Dockery} are that for any well-behaved $ \alpha(x)$  the \emph{slow} species \emph{always} drives the fast one to extinction, which is  quite unlike the results of Figure 1(b). This result is independent of initial conditions and the detailed form of  $\alpha$. The qualitative reason for this result is that F cannot exploit the oases as well as S since it is always wandering into the deserts. However, our results show that even for reasonably large values of $f$ and $s$ effects left out of the partial differential equations (e.g population fluctuations) change the results in a qualitative way. For very large $N_o$, we  recover the continuum results.
 
 Actual populations could well be in the regime where the individual-based model differs from the continuum one, so it is useful to enquire about the sources of the complex behaviour.  Situations where reaction-diffusion equations fail to capture qualitative features of stochastic dynamics are well known \citep{Witten, Derrida, KSN, Shnerb,Evolution}. For example, in the DLA model \citep{Witten, Sander}, fluctuations in growth rate drive instabilities and produce fractal shapes. For  Fisher-KPP dynamics \citep{Fisher,Kolmogorov} the discreteness of agents in an agent-based model dominates front propagation \citep{Derrida, KSN, Panja}. Discreteness also can qualitatively change low-dimensional dynamical models \citep{King}.  In our case both fluctuation and discreteness effects are important.
 
 First consider the case of large $\eta$. Here the dominant effect is the cost of migration in crossing the desert in our landscape. The reaction-diffusion equations, above, predict an exponentially small tail of the slow species surviving into the fast oasis which eventually starts growth there. For discrete agents, however, this is an overestimate: we should not consider values of $f$ which correspond to far less than one agent. Thus the tail is suppressed, and the number of agents that can cross the desert is very small, and
 are easily killed off by fluctuations. We can qualitatively represent this effect at the continuum level \citep{Derrida}  by changing the birth term in the reaction-diffusion equations:
 \begin{equation}
 \label{cutoff}
 \alpha(x)  f \to \alpha(x) f \Theta(f - \epsilon) \quad \quad
\alpha(x) s \to \alpha(x)  s \Theta(s - \epsilon),
\label{cutoffeqn}
\end{equation}
where the function $\Theta$ is 0 for negative arguments and 1 for positive, and $\epsilon$ is a parameter of order unity. This means that when the density drops to very low values  (less than one individual in a few lattice sites) , we do not expect to have reproduction. Numerical solutions of the new reaction-diffusion equations are shown in Figure 2. Now we have a region of stable coexistence. This is qualitatively similar to the right side of Figure 1(b). In the full agent-based model, the coexistence region is metastable: after a long time the oasis of the slow agents is invaded by the fast whose transition across the desert is easier due to the large value of $D_f$. 
\begin{figure}
\includegraphics[width=2.8in]{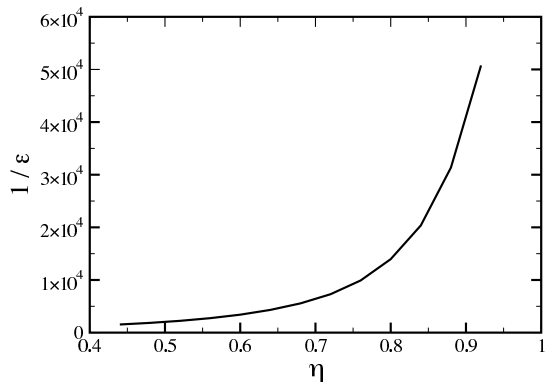}
\caption{
The phase boundary marking the critical value of $1/\epsilon$ (i.e., the critical value of $N_o$ for $\epsilon=1$) as a function of $\eta$ which separates  the coexistence phase (below the line) and
the victory of the slow species (above the line). Note the  qualitative similarity to Figure 1b, right side.}
\end{figure}

The role of discreteness and fluctuations for small $\eta$, i.e. a nearly uniform environment, is more subtle, and is a separate mechanism. We will argue that density fluctuations increase the death rate: since F smooths out fluctuations by diffusion, S is at a disadvantage. The basic reason for this is that the birth rate is linear in our model and the death rate non-linear. Consider the case of $\alpha$ constant, and suppose  the density of the slow species is given by $s(x)=s_o + s_k \cos(kx)$, where the $k$'s are chosen to satisfy no-flux boundary conditions. The $s_k$ are the amplitudes of the fluctuations of various wavelengths. Now it is clear that the total number of births in time $dt$  is independent of the fluctuations: 
\begin{equation}
dt  \int \alpha s dx = dt \alpha L s_o.
\label{births}
\end{equation}
However the total number of deaths is:
\begin{equation}
dt \int \beta  s^2 dx= dt \beta L (s_o^2+\sum s_k^2 /2).
\label{deaths}
\end{equation}
Fluctuations increase the death rate. As we have noted, we expect the fluctuations of $f(x)$ to be smaller, so the slow species is driven to extinction.



To treat this effect in more detail we should write down the master equation for the stochastic process described above. Then we could find the differential equations for the averages of $f,s$ as moments.  The reaction-diffusion equations, Eq. (\ref{RD})  should be thought of as the mean-field decoupling of these equations. 

For our purposes it is sufficient to average over position as well as over internal fluctuations. We denote  these processes by  $\langle \cdot \rangle$. The diffusion terms average to zero, and the remaining terms in the moment equations must be of the form:
\begin{eqnarray}
 \langle \dot f \rangle & = & \langle \alpha f \rangle- \beta(\langle f^2 \rangle+ \langle fs \rangle)
 \nonumber \\
 \langle \dot s \rangle & = &  \langle \alpha s \rangle- \beta(\langle s^2 \rangle+ \langle fs \rangle).
 \label{aveRD}
 \end{eqnarray}
Now we can subtract the two equations and find:
\begin{eqnarray}
\label{correqn}
\langle \dot f - \dot s \rangle  &=&    \langle f-s \rangle [ \langle \alpha \rangle - 
 \beta \langle  f+s \rangle ] 
\nonumber \\
& + & [\mathrm{Cov}(\alpha,f)  -  \mathrm{Cov}(\alpha,s)] \nonumber \\
&-& \beta[\mathrm{Var}(f) - \mathrm{Var}(s)].
\end{eqnarray} 
Here $\mathrm{Cov}(\alpha,f) =  \langle \alpha f \rangle -  \langle \alpha \rangle  \langle f \rangle$ is the covariance, and $\mathrm{Var}(f) =  \langle f^2 \rangle - \langle f \rangle^2$ is the variance.   

\begin{figure}
\begin{center}
\includegraphics[width=3in]{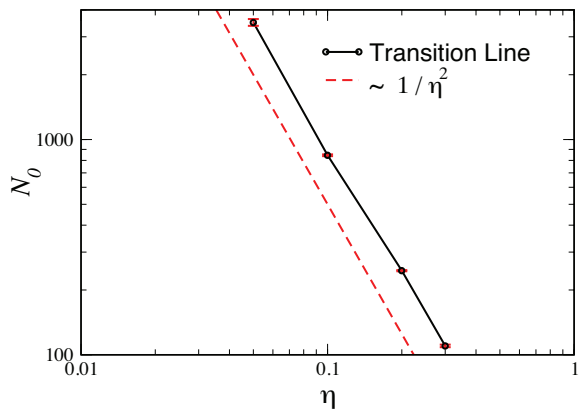}
\caption{A comparison of the division line between fast and slow dominance (the left side of Figure 1b) with the estimate based on Eq. (\ref{correqn}). The simulations are for a system of length 10.}
\label{scaling}
\end{center}
\end{figure}

Suppose $f$ and $s$ are very close to one another. Then the first term in the equation is zero. The second term favours the slow species since it can better  follow the variation of $\alpha$ as in the deterministic treatment \citep{Dockery}. The third term favours the fast species since its variance is smaller. We can estimate the second term by setting  either density to be  of order $N_o \alpha$. The difference in the covariances is of order $N_o  \langle \alpha^2 \rangle \propto N_o \eta^2$. In the last term the variances are of order $N_o$ so that $\beta [ \mathrm{Var}(f)- \mathrm{Var}(s)]$ is of order unity. The two terms are equal on the division line on Figure 1(b), so that  $N_o \propto 1/\eta^2$ defines the  line. This estimate is verified by our results, see Figure \ref{scaling}.

An example of these mechanisms in a more realistic context may be found in \citep{Hanski}. This is a spatially explicit study of the evolution of dispersal rates in a patchy environment which is closely based on the ecology of the checkerspot butterfly. The model is quite complex, and has 15 parameters that are deduced from field observations.  In the study there is a cost of migration between patches of habitat, and the dispersal rate is allowed to evolve. As the cost of migration increases, dispersal rates decrease, but for large costs (analogous to large $\eta$ in our treatment) there is non-monotonic behaviour, and the dispersal rate increases again. We propose that the underlying mechanisms in the model are revealed in our schematic treatment, where the evolutionary result of one or another species going extinct is a proxy for evolution of dispersal rate. We think that this kind of phenomenon, where fluctuation-driven effects that occur at relatively low densities may be operative in real systems, and will repay further study. 
In particular, an experimental realization of the model is probably possible. For example, In a predator-prey system of bacteria \cite{Holyoak} fluctuation effects on extinction have been demonstrated in dispersed environments. It would be very interesting to do a similar study in the setting of competition for 
resources.

Finally we return to the question of evolution in our model. In \cite{HM,CHM,Hanski} there is an optimal dispersal rate. We have numerical evidence that this is also true for our model. That is, for each $N_o$ and $\eta$ there is a $\bar D(N_o,\eta)$ which wins any competition and presumably is the evolutionarily stable state. For example, we find that $\bar D(100,0.75) \approx 0.25$ and $\bar D(1000,0.75) \approx 0.15$. The results of \cite{Dockery} may be interpreted to mean that $\lim_{N_o  \to \infty} \bar D(N_o, \eta) =0$ for all $\eta$.

LMS is supported in part by National Science Foundation grant DMS 0553487. DAK is supported in part by the Israel Science Foundation. We would like to thank Nadav Shnerb, Evgeniy Khain, Glenn Strycker, Aaron King, Charlie Doering, and Jack Waddell for useful discussions. 


\begin{thebibliography}{16}
\expandafter\ifx\csname natexlab\endcsname\relax\def\natexlab#1{#1}\fi
\expandafter\ifx\csname bibnamefont\endcsname\relax
  \def\bibnamefont#1{#1}\fi
\expandafter\ifx\csname bibfnamefont\endcsname\relax
  \def\bibfnamefont#1{#1}\fi
\expandafter\ifx\csname citenamefont\endcsname\relax
  \def\citenamefont#1{#1}\fi
\expandafter\ifx\csname url\endcsname\relax
  \def\url#1{\texttt{#1}}\fi
\expandafter\ifx\csname urlprefix\endcsname\relax\def\urlprefix{URL }\fi
\providecommand{\bibinfo}[2]{#2}
\providecommand{\eprint}[2][]{\url{#2}}

  
  \bibitem[{\citenamefont{Johnson and Gaines}(1990)}]{Johnson}
\bibinfo{author}{\bibfnamefont{M.~L.} \bibnamefont{Johnson}} \bibnamefont{and}
  \bibinfo{author}{\bibfnamefont{M.~S.} \bibnamefont{Gaines}},
  \bibinfo{journal}{Annual Review of Ecology and Systematics}
  \textbf{\bibinfo{volume}{21}}, \bibinfo{pages}{449} (\bibinfo{year}{1990}).
  
  \bibitem[{\citenamefont{Hastings}(1983)}]{Hastings}
\bibinfo{author}{\bibfnamefont{A.}~\bibnamefont{Hastings}},
  \bibinfo{journal}{Theoretical Population Biology}
  \textbf{\bibinfo{volume}{24}}, \bibinfo{pages}{244} (\bibinfo{year}{1983}).
  
  \bibitem[{\citenamefont{Dockery et~al.}(1998)\citenamefont{Dockery, Hutson,
  Mischaikow, and Pernarowski}}]{Dockery}
\bibinfo{author}{\bibfnamefont{J.}~\bibnamefont{Dockery}},
  \bibinfo{author}{\bibfnamefont{V.}~\bibnamefont{Hutson}},
  \bibinfo{author}{\bibfnamefont{K.}~\bibnamefont{Mischaikow}},
  \bibnamefont{and}
  \bibinfo{author}{\bibfnamefont{M.}~\bibnamefont{Pernarowski}},
  \bibinfo{journal}{J. Math. Biology} \textbf{\bibinfo{volume}{37}},
  \bibinfo{pages}{61} (\bibinfo{year}{1998}).


\bibitem{HM} W. D. Hamilton and R. M. May, Nature {\bf 269}, 578 (1977)

\bibitem{CHM} H. N. Comins, W. D. Hamilton, and R. M. May, J. Theo. Biol. {\bf 82}, 205 (1980).


\bibitem[{\citenamefont{Brunet and Derrida}(1997)}]{Derrida}
\bibinfo{author}{\bibfnamefont{E.}~\bibnamefont{Brunet}} \bibnamefont{and}
  \bibinfo{author}{\bibfnamefont{B.}~\bibnamefont{Derrida}},
  \bibinfo{journal}{Phys. Rev. E} \textbf{\bibinfo{volume}{56}}
  (\bibinfo{year}{1997}).

\bibitem[{\citenamefont{Kessler et~al.}(1998)\citenamefont{Kessler, Ner, and
  Sander}}]{KSN}
\bibinfo{author}{\bibfnamefont{D.~A.} \bibnamefont{Kessler}},
  \bibinfo{author}{\bibfnamefont{Z.}~\bibnamefont{Ner}}, \bibnamefont{and}
  \bibinfo{author}{\bibfnamefont{L.~M.} \bibnamefont{Sander}},
  \bibinfo{journal}{Phys. Rev. E} \textbf{\bibinfo{volume}{58}},
  \bibinfo{pages}{107} (\bibinfo{year}{1998}).

\bibitem[{\citenamefont{Witten and Sander}(1981)}]{Witten}
\bibinfo{author}{\bibfnamefont{T.~A.} \bibnamefont{Witten}} \bibnamefont{and}
  \bibinfo{author}{\bibfnamefont{L.~M.} \bibnamefont{Sander}},
  \bibinfo{journal}{Phys. Rev. Lett.} \textbf{\bibinfo{volume}{47}},
  \bibinfo{pages}{1400} (\bibinfo{year}{1981}).

\bibitem[{\citenamefont{Heino and Hanski}(2001)}]{Hanski}
\bibinfo{author}{\bibfnamefont{M.}~\bibnamefont{Heino}} \bibnamefont{and}
  \bibinfo{author}{\bibfnamefont{I.}~\bibnamefont{Hanski}},
  \bibinfo{journal}{American Naturalist} \textbf{\bibinfo{volume}{157}},
  \bibinfo{pages}{495} (\bibinfo{year}{2001}).

\bibitem[{\citenamefont{Shnerb et~al.}(2000)\citenamefont{Shnerb, Louzon,
  Bettelheim, and Solomon}}]{Shnerb}
\bibinfo{author}{\bibfnamefont{N.~M.} \bibnamefont{Shnerb}},
  \bibinfo{author}{\bibfnamefont{Y.}~\bibnamefont{Louzon}},
  \bibinfo{author}{\bibfnamefont{E.}~\bibnamefont{Bettelheim}},
  \bibnamefont{and} \bibinfo{author}{\bibfnamefont{S.}~\bibnamefont{Solomon}},
  \bibinfo{journal}{PNAS} \textbf{\bibinfo{volume}{97}}, \bibinfo{pages}{10322}
  (\bibinfo{year}{2000}).

\bibitem[{\citenamefont{Tsimring et~al.}(1996)\citenamefont{Tsimring, Levine,
  and Kessler}}]{Evolution}
\bibinfo{author}{\bibfnamefont{L.~S.} \bibnamefont{Tsimring}},
  \bibinfo{author}{\bibfnamefont{H.}~\bibnamefont{Levine}}, \bibnamefont{and}
  \bibinfo{author}{\bibfnamefont{D.~A.} \bibnamefont{Kessler}},
  \bibinfo{journal}{Phys. Rev. Lett.} \textbf{\bibinfo{volume}{76}},
  \bibinfo{pages}{4440} (\bibinfo{year}{1996}).

\bibitem[{\citenamefont{Sander}(2000)}]{Sander}
\bibinfo{author}{\bibfnamefont{L.~M.} \bibnamefont{Sander}},
  \bibinfo{journal}{Contemporary Physics} \textbf{\bibinfo{volume}{41}},
  \bibinfo{pages}{203} (\bibinfo{year}{2000}).

\bibitem[{\citenamefont{Fisher}(1937)}]{Fisher}
\bibinfo{author}{\bibfnamefont{R.}~\bibnamefont{Fisher}},
  \bibinfo{journal}{Annals of Eugenics} \textbf{\bibinfo{volume}{7}},
  \bibinfo{pages}{355} (\bibinfo{year}{1937}).

\bibitem[{\citenamefont{Kolmogorov et~al.}(1937)\citenamefont{Kolmogorov,
  Petrovsky, and Piscounov}}]{Kolmogorov}
\bibinfo{author}{\bibfnamefont{A.~I.} \bibnamefont{Kolmogorov}},
  \bibinfo{author}{\bibfnamefont{I.}~\bibnamefont{Petrovsky}},
  \bibnamefont{and}
  \bibinfo{author}{\bibfnamefont{N.}~\bibnamefont{Piscounov}},
  \bibinfo{journal}{Moscow Univ. Bull Math. A} \textbf{\bibinfo{volume}{1}},
  \bibinfo{pages}{1} (\bibinfo{year}{1937}).

\bibitem[{\citenamefont{Panja}(2004)}]{Panja}
\bibinfo{author}{\bibfnamefont{D.}~\bibnamefont{Panja}},
  \bibinfo{journal}{Physics Reports} \textbf{\bibinfo{volume}{393}},
  \bibinfo{pages}{87} (\bibinfo{year}{2004}).

\bibitem[{\citenamefont{Henson et~al.}(2001)\citenamefont{Henson, Costantino,
  Cushing, Desharnais, Dennis, and King}}]{King}
\bibinfo{author}{\bibfnamefont{S.~M.} \bibnamefont{Henson}},
  \bibinfo{author}{\bibfnamefont{R.~F.} \bibnamefont{Costantino}},
  \bibinfo{author}{\bibfnamefont{J.~M.} \bibnamefont{Cushing}},
  \bibinfo{author}{\bibfnamefont{R.~A.} \bibnamefont{Desharnais}},
  \bibinfo{author}{\bibfnamefont{B.}~\bibnamefont{Dennis}}, \bibnamefont{and}
  \bibinfo{author}{\bibfnamefont{A.~A.} \bibnamefont{King}},
  \bibinfo{journal}{Science} \textbf{\bibinfo{volume}{294}},
  \bibinfo{pages}{602} (\bibinfo{year}{2001}).

%

  
  \bibitem{Holyoak} M. Holyoak and S. P. Lawler, Ecology {\bf 77}, 1867 (1996). 

\end{thebibliography}

\end{document}